\documentclass[10pt]{article}
\renewcommand{\a}{{\bf a}}

\renewcommand{\b}{{\bf b}}
\renewcommand{\c}{{\bf c}}
\makeatletter \renewcommand\@biblabel[1]{$#1.$} \makeatother

\newcommand{\K}{{\cal K}}

\newcommand{\EQ}{\begin{equation}}
\newcommand{\EE}{\end{equation}}

\usepackage{amsmath}
\usepackage{amssymb}
\usepackage{graphicx}
\usepackage{cite}
\usepackage{color} 
\usepackage{setspace} 
\usepackage[labelfont=bf,labelsep=period,justification=raggedright]{caption}

\makeatletter
\renewcommand{\@biblabel}[1]{\quad#1.}
\makeatother

\date{}

\begin{document}

\begin{flushleft}
{\Large
\textbf{Formation of regulatory modules by local sequence duplication}
}
\\
Armita Nourmohammad$^{1}$, 
 Michael L\"assig$^{1}$
\\
\bf{1} Institute for Theoretical Physics, University of Cologne, Z\"ulpicherstr.~77, 50937 K\"oln, Germany
\\

\end{flushleft}

\section*{Abstract}

Turnover of regulatory sequence and function is an important part of molecular evolution. But what are the modes of sequence evolution leading to rapid formation and loss of regulatory sites? Here, we show that a large fraction of neighboring transcription factor binding sites in the fly genome have formed from a common sequence origin by local duplications. This mode of evolution is found to produce regulatory information: duplications can seed new sites in the neighborhood of existing sites. Duplicate seeds evolve subsequently by point mutations, often towards binding a different factor than their ancestral neighbor sites. These results are based on a statistical analysis of 346 cis-regulatory modules in the {\em Drosophila melanogaster} genome, and a comparison set of intergenic regulatory sequence in {\em Saccharomyces cerevisiae}. In fly regulatory modules, pairs of binding sites show significantly enhanced sequence similarity up to distances of about $50$ bp. We analyze these data in terms of an evolutionary model with two distinct modes of site formation: (i)~evolution from independent sequence origin and (ii)~divergent evolution following duplication of a common ancestor sequence.  Our results suggest that pervasive formation of binding sites by local sequence duplications distinguishes the complex regulatory architecture of higher eukaryotes from the simpler architecture of unicellular organisms.

\section*{Introduction}
The importance of regulatory variations as a driving force for phenotypic evolution has been suggested some time ago~\cite{monod, KW}. However, a quantitative understanding of gene regulation has become possible only after the advent of large-scale genomic sequence and regulatory interaction data.  Important building blocks are genome-wide maps of protein-DNA binding, statistical inference methods ~\cite{BV,SF}, high-throughput measurements of sequence-specific binding affinities of transcription factors ~\cite{FHAS,MBJWMSYB,MQ,BBPTGJ}, and cross-species comparisons of regulatory sequences and regulatory functions~\cite{STARK}. 

The resulting picture is quite diverse: Core parts of developmental regulatory networks can be conserved over large evolutionary ranges~\cite{DE}, and individual promoters in flies are found to be conserved in function over large evolutionary distances ~\cite{LBPK,HPIME}. Functional changes in promoters have been identified as well, but the relative roles of adaptive and near-neutral evolution remain to be clarified. The sequences in regulatory DNA regions evolve under less constraint  than their functional output. This feature can be explained by wide-spread compensatory changes, which have been identified between different nucleotides within individual binding sites as well as between different sites within a promoter~\cite{MKCL,LK,LPK,LBPK,LPABNK,AK,HPIME}. At the promoter level, this dynamics includes loss and gain of binding sites, the rates of which have been estimated in flies and yeast~\cite{MPNILBE, DF,MKCL}. The observed site turnover is consistent with moderate negative selection acting on individual sites~\cite{MCPIE,MKCL}, whereas the function of entire promoters is under stronger stabilizing selection~\cite{LBPK}. 

The evolutionary constraint of regulatory sequence and function depends on the level of complexity in promoter architecture. Prokaryotes and unicellular eukaryotes have short intergenic regions, and regulatory functions are often encoded by few binding sites. The more complex cis-regulatory information in higher eukaryotes is organized into {\em regulatory modules}, which are typically a few hundred base pairs long and are spatially separated by larger segments of intergenic DNA~\cite{OGH,BPRHG}. Within modules, regulatory functions often depend on clusters of neighboring binding sites for multiple transcription factors, which are coupled by cooperative interaction~\cite{Ptashne,SSUGS,HGLRMD,DAVIDSON,LYNCH}. Bioinformatic algorithms trace such site clusters to identify regulatory DNA regions \cite{RVGS,HGCM,SNS,BPLSREC,ABWG ,LZHSH,LE2009}. The relative order and spacing of sites within clusters follows a regulatory ``grammar'', which distinguishes functionally neutral site changes from rearrangements affecting promoter function~\cite{SSL,SAL,MMML,KA,AK,LE2010}. 
The combinatorial complexity of this grammar ensures the specificity of regulation in the larger genomes of multicellular eukaryotes~\cite{BGH,LT}. At the same time, the grammar is flexible enough to allow substantial sequence evolution in a regulatory module while maintaining its overall functional output. 

In addition to point mutations, sequence insertions and deletions (indels) play a significant role in this dynamics. Several studies have noted the prevalence of repetitive sequence elements in promoter regions and their potential influence on regulatory function~\cite{BRITTEN,HSBD,VLCHV,BRPM,TS,MA,SS,Gruen}. In particular, a recent detailed analysis of the evolutionary rates of short tandem repeats in {\em Drosophila} has shown a net surplus of insertions, suggesting that these repeats may produce new regulatory sequence~\cite{SS}. But to what extent is this actually the case? A priori, the link between repeat evolution and regulation is far from obvious: Duplications in repeats can either be part of the neutral background evolution in regulatory sequences, or increase the spacing between existing binding sites of a regulatory module, or contribute to the formation of new sites. Disentangling these roles is subtle, because detected tandem repeats in contemporary sequence overlap with only a small fraction of binding sites, motif size and total length of most repeats are shorter than length and spacing of typical binding sites in a cluster, and repeat lifetimes are much shorter than conservation times of regulatory elements~\cite{Gruen}. Hence, the  role of repeat dynamics for regulation is an open problem: Do local duplications actually transport and produce regulatory information? 

This is the topic of the present paper. We show that local duplications have left a striking signature in the fly genome: the majority of transcription factor binding sites in regulatory modules show evidence of a duplication event in their evolutionary history. We conclude that over long evolutionary times, local duplications are pervasive and crucial for the formation of complex regulatory modules in the fly genome. This mode of evolution sets the speed of regulatory evolution  and facilitates adaptive changes of promoter function. We infer site duplications from their traces in the sequence of neighboring binding sites, but most duplication events predate the tandem repeats present in contemporary sequence. This distinguishes our study from comparative analysis of regulatory sequence between closely related species~\cite{SS,TS,Gruen,MA, BRPM}, which can detect the insertion-deletion dynamics of contemporary repeats, but cover only a small window in the evolution of regulatory sites.

The importance of binding site evolution by duplication is grounded in the biophysics of transcription factor-DNA interactions: the sequence-dependent probability of binding between factor and site depends in a strongly nonlinear way on the binding energy~\cite{BV}: it takes values close to 1 in an energy range below the maximum binding energy, then drops rapidly as the energy decreases further, and is close to 0 in the energy range of non-binding sites.  This nonlinearity generates strong epistatic effects for point mutations within binding sites~\cite{BWL,MKCL} and, in turn, an asymmetry in the turnover of binding sites.  Functional sites can rapidly lose their binding affinity to a factor by one or two point mutations. Rapid adaptive formation of a site, however, requires a {\em seed sequence} with marginal binding, to which positive selection for point mutations towards stronger binding can latch on. Such seeds are contained in random sequence, but at unspecific positions. Estimates of the rate of site formation based on biophysically grounded fitness models suggest that point mutations alone can explain the rapid formation of an individual site in a sufficiently large sequence interval, but not the formation of spatially confined agglomerations of sites characteristic of regulatory modules~\cite{BL,BWL,LASSIG}. As we show in this paper, local sequence duplications generate seeds for new sites specifically in the neighborhood of functional sites. 

Our analysis proceeds in three steps. First, we analyze local sequence similarities in regulatory regions of the {\em Drosophila melanogaster} genome in a model-independent way. In regulatory modules, we find a  significant autocorrelation in nucleotide content for distances up to about 70 bp. This autocorrelation includes the known contributions of tandem repeat sequences, but it extends to a much larger distance range. The signal turns out to be generated by local sequence clusters, a substantial fraction of which are functional transcription factor binding sites with similar sequence motifs. 
 In the second part of the paper, we infer the evolutionary origin of these correlated pairs of binding sites, using a probabilistic model with mutations, genetic drift, and selection. 
The model compares the likelihood of two alternative histories: a pair of sites evolves either independently or by duplication from a common ancestor sequence. The duplication is followed by diversification under selection for binding of two (in general different) factors. We show that the duplication pathway is the most likely history for pairs of sites with a mutual distance up to about 50 bp, and we find evidence that this pathway   is specific to  regulatory modules of multicellular eukaryotes. Finally, we show that the duplication mode has adaptive potential: duplicated ancestor sites can act as seeds for the subsequent formation of a novel binding site for the same factor and, notably, even for a different factor.

\section*{Results}

\subsection*{Sequence autocorrelation in regulatory DNA}
The most straightforward measure of local similarity in a sequence segment is the {\em autocorrelation function}, which is defined as the difference between the likelihood $c(r)$ that two nucleotides at a distance of $r$ base pairs are identical and mean identity $c_0$ of two random nucleotides, $\Delta (r) = c (r) - c_0$. This function is straightforward to evaluate from sequence data as given by eq.~(\ref{Delta}) in Materials and Methods.   We have obtained the autocorrelation function in 346 regulatory modules of the {\em D. melanogaster} genome with  length of more than $1000$ bp identified by REDfly database \cite{BCC,GLHH,HGB}. The results are shown in Fig.~\ref{fig:Infoauto} (a). 
In the distance range up to about 70 bp, the function $\Delta (r)$ takes positive values that decay with $r$ in a roughly exponential way; this signal is clearly above the noise level.  The mean identity is evaluated in a local window of 500 bp (changing the window length affects the baseline of this function, but not its short-distance behavior). The autocorrelation signal is small and has several potential sources, such as multiple binding sites for similar motifs, microsatellite and minisatellite repeats at short length scales~\cite{SS,MA,Gruen,TS}, homopolymeric stretches of nucleotides characteristic of nucleosome-depleted regions~\cite{SW}, or other local inhomogeneities in sequence composition.
As a next step, we will characterize local sequence similarity in a more specific way: we will show that mutually correlated nucleotide pairs are not evenly distributed over regulatory modules, but occur in local clusters with a characteristic length scale of around 7 bp. This signal will be analyzed from an evolutionary point of view and be linked to cis-regulatory function.

\subsection*{Sequence motifs and information}

 To motivate the following analysis, assume that a given sequence segment is covered by families of sites belonging to different {\em motifs}. By definition, a motif  is a probability distribution $Q(\a)$ of genotypes $\a = (a_1, \dots, a_\ell)$, which describes a specific set of sequence sites with $\ell$ consecutive base pairs and is different from the background distribution $P_0(\a)$. The statistical deviation of a motif from background is measured by the relative entropy between these distributions, $H(Q|P_0)$, which is given by eq.~(\ref{H}) in Materials and Methods. This quantity determines the average {\em sequence information} per site, which is often quoted in units of bits~\cite{SF}. Multiplying $H(Q | P_0)$ with the number of sites for each motif and summing over all motifs produces a measure of the total sequence information contained in a genomic region.

Well-known motifs in regulatory DNA are the families of binding sites for a given transcription factor. In eukaryotic systems, these sites have a typical length of about 5-10 bp and frequency distributions $Q$ (called position weight matrices) with a typical information content $H \approx 6-8$ bits per site; see the recent discussion by ~\cite{WM}. Other motifs can be defined, for example, in nucleosome-depleted sequences in eukaryotes and for repeat units in tandem repeats. If all motifs occurring in a given sequence segment were known, we could try to predict their sites and evaluate the information content directly. In the present part of the analysis, we proceed differently. 
We only assume that sequence motifs carry a certain information content over sites of a given length $\ell$, but we make no further assumptions on position weight matrices, sequence coverage, or evolutionary origin. We can still recover part of this sequence information from those motifs that occur more than once in the sequence segment. A pair of sites of length $\ell$ belonging to the same motif has an average {\em similarity information} given by the mutual entropy $K (c, \ell |c_0)$, which measures the enhanced similarity $c$ of aligned nucleotides of the site sequences compared to the background similarity $c_0$ and is given by eq.~(\ref{K}) in Materials and Methods. Clearly, the similarity information between pairs of sites is a somewhat diluted measure of the full information content due to motifs. As a rule of thumb, the mutual entropy per site pair, $K(c, \ell | c_0)$, recovers about half of the sequence information per site,  $H(Q| P_0)$. For example, binding sites for the same transcription factor are strongly correlated, with a typical similarity $c \approx 0.7$ and a similarity information $K \approx3 $ bits per site pair. 

Here, we want to identify pairs of similar sites at a given distance $r$ and relate them to the sequence autocorrelation function $\Delta (r)$ discussed above. Thus, we estimate the total similarity information $\K_\ell (r)$ per unit sequence length of all strongly correlated pairs of sites with distance $r$ and length $\ell$ in regulatory modules. This quantity can be defined by constructing a set of site pairs for given $r$ and $\ell$ with the following properties: (i)~Any pair of sites has an average mutual similarity between aligned nucleotides above a certain threshold, $c > c_{\min} (\ell)$. (ii)~The left sites (and, hence, also the right sites) of all pairs have no mutual overlaps. This condition is necessary in order to avoid overcounting of mutual similarity in overlapping site pairs. (iii)~The sum of the mutual similarities of all pairs in the set is maximal (see Fig.~\ref{fig:frameshift} for illustration). This condition is also used to set the similarity threshold $c_{\min} (\ell)$. To identify a set of site pairs with properties~(i) to~(iii), we use a dynamic programming algorithm as explained in Materials and Methods. This method allows for optimization of sequence length $\ell$ similar to the procedure in local sequence alignment algorithms~\cite{DEKM}. In the maximum-similarity set, we record the average mutual similarity $\bar c (r,\ell)$ of aligned nucleotides in site pairs, which determines the mean information content per site pair, $K(\bar c(r,\ell), \ell | c_0)$ (see eq.~(\ref{K}) in Materials and Methods). We also record the number $n(r,\ell)$ of site pairs and determine the excess  $\Delta n(r,\ell) = n(r,\ell) - n_0 (\ell)$ over the number expected by chance in background sequence, $n_0 (\ell)$ (see Materials and Methods). The distance-dependent total similarity information per unit length in a  sequence segment of size $L$ can then be estimated as $\K_\ell (r) = (\Delta n(r,\ell) /L) \, K(\bar c(r,\ell), \ell | c_0) $. 

Our inference of $\K_\ell (r)$ is related to recent methods for prediction of unknown regulatory modules based on their enhanced sequence similarity contained in words of length $\ell$~\cite{ABWG ,LZHSH,LE2009}. But the evaluation of sequence similarity and the goals of the analysis differ: module prediction uses the total similarity in a genomic region, which in our setup is given by summation of $\K_\ell (r)$ over all distances $r$ and over different word lengths $\ell$. Our analysis is limited to known regulatory modules and focuses on the dependence of $\K_\ell (r)$ on $r$ and $\ell$. A specific part of this signal, obtained from sites with distance $r$ below  50 bp, will be associated below with local duplications as prevalent evolutionary mode.

\subsection*{Similarity information in regulatory modules of  {\em Drosophila}}

\begin{figure}[!ht]
\raggedright{}\includegraphics[scale=0.47]{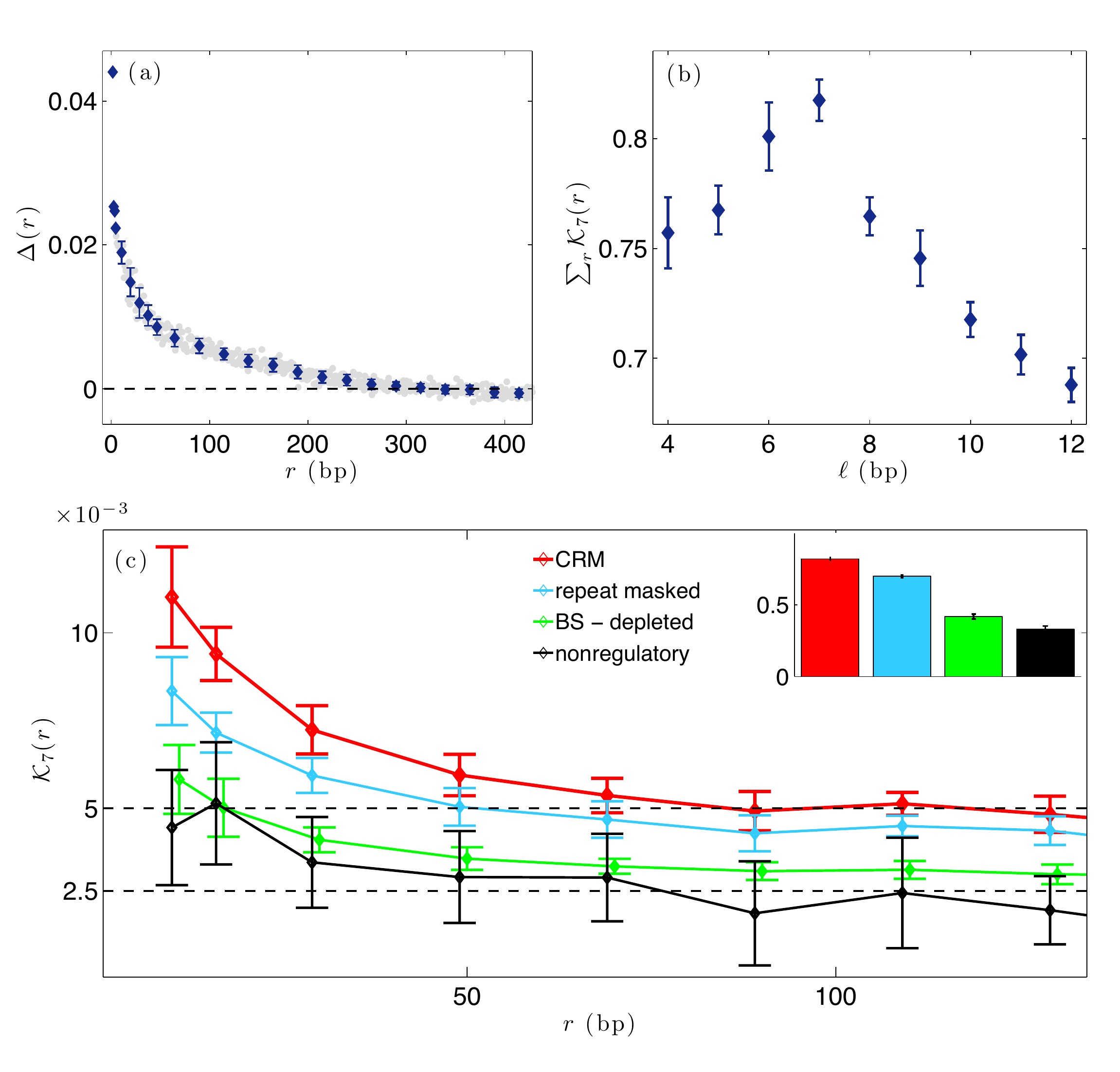}
\par
\caption{\label{fig:Infoauto} {\small {\bf Sequence similarity in regulatory modules of the fly genome.}
(a)~Sequence autocorrelation $\Delta(r)$ as a function of distance $r$, obtained from $346$ regulatory modules in \textit{D. melanogaster} (gray: unbinned data, blue: binned in intervals of variable length).  The autocorrelation values are positive and depend on $r$ in a roughly exponential way up to about $70$ bp.  
(b)~Total similarity information  $\K_{\rm tot} (\ell)=\sum_{r=1}^{100} \K_{\ell}(r)$ as a function of motif length $\ell$ for all pairs of strongly correlated sites with mutual distance $r < 100$ bp in the same set of regulatory modules. This function takes its maximum at a characteristic motif length of $\ell = 7$ bp. (c)~Distance-dependent similarity information $\K_{7} (r)$ for motif length $\ell = 7$ evaluated in all sequence (red), binding site-masked sequence (green), repeat-masked sequence (blue) in regulatory modules, and in generic intergenic sequence (black). Repeat-masked sequence is generated using the Tandem Repeat Finder~\cite{BENSON} with match-mismatch-indel penalty parameters (2,3,5). Insert: Total similarity information $\K_{\rm tot} (\ell \! = \! 7)$ for the same sequence categories. Binding sites, but not tandem repeats, account for a substantial fraction of the similarity information.}}
\end{figure}

 We evaluate the similarity information in the set of 346 regulatory modules of {\em Drosophila melanogaster} and in surrounding background sequence. The following features of local sequence similarity can be extracted:

\noindent
---{\em The total information of local sequence similarity is maximal for motifs of length $\ell = 7$.} Fig.~\ref{fig:Infoauto}(b) shows the total similarity information of all detected site pairs in the range of up to 100 bp, $\K_{\rm tot} (\ell) = \sum_{r = \ell}^{100} \K_\ell (r)$, as a function of the site length $\ell$. The function $\K_{\rm tot} (\ell)$ takes its maximum, that is, the similarity information is most significant, for $\ell = 7$. The signal falls off at shorter length scales, because typical motif sequences are only partially covered, and at larger length scales, because uncorrelated flanking nucleotides contribute negatively to the similarity information. In this sense, detected motifs cover a characteristic length of about 7 bp. A similar length scale has been observed in tandem repeats~\cite{BRPM,MA,TS}. 

\noindent 
---{\em The function $\K_7 (r)$ takes distance-dependent positive values in the range of up to 50 bp and saturates to a positive asymptotic value for larger distances.}  Thus, its distance dependence is compatible to that of the sequence autocorrelation function $\Delta (r)$ shown in Fig.~\ref{fig:Infoauto}(a). This pattern is due to site pairs with high mutual similarity,  $c > 0.85$.

\noindent
---{\em Correlated binding sites explain a substantial part of the similarity information.} We estimate this contribution by masking all functional sites~\cite{HGB,GLHH,BCC} and re-evaluating the function $\K_7 (r)$ in their sequence complement; see Fig.~\ref{fig:Infoauto}(c). Known binding sites cover about $10\%$ of the regulatory modules, but the signal is reduced by about 50\%, indicating that these sites are an important source of similarity information. The binding site-masked signal is comparable to its counterpart  $\K_7 (r)$ in non-regulatory intergenic sequence. 

 \noindent
---{\em Microsatellite repeats explain only a small part of the similarity information.} 
We identify such repeats using the Tandem Repeat Finder~\cite{BENSON}. If we remove about 5 \% of the sequence in regulatory modules as repeats, the similarity information is reduced by less than 10\%; Fig.~\ref{fig:Infoauto}(c). This is not surprising, because our sequence similarity measure differs from that of repeat analysis. In particular, our measure is sensitive to correlated segments on larger distance scales than typical tandem repeats, because it does not require a contiguous interval of self-similar sequence in between.  

\noindent
---{\em Homologous regions in other fly genomes show a consistent form of  $\K_7 (r)$.}
We analyze homologous regions of two other {\em Drosophila} species,  {\em D. yakuba} and {\em D. pseudoobscura} (see Materials and Methods). 
As shown in Fig.~\ref{fig:crossSpec}, these putative regulatory modules have patterns $\K_7 (r)$ of very similar overall amplitude and distance-dependence, with enhanced values in the range of up to 50 bp.  

In summary, our model-independent analysis shows that  motifs with a characteristic length of about 7 bp play an important part in the distance-dependent sequence autocorrelation of {\em Drosophila} regulatory modules. The characteristic length coincides with the typical length of binding sites, and a substantial fraction of the signal can be explained by sequence correlations involving known binding sites. Therefore, we now focus the analysis on this smaller, but experimentally validated set of  sites~\cite{BCC,GLHH,HGB}  and analyze in detail the evolutionary mechanism generating the sequences of neighboring pairs of sites.

\subsection*{Evolutionary modes of binding sites}
Binding sites are ideal objects to study the production of information by sequence evolution. The sequence motif is approximately known for about $70$ transcription factors in {\em Drosophila}, that is, we can analyze the full position-dependent sequence information of these motifs, not just the similarity information of motif pairs. Furthermore, there is a simple link between sequence statistics and evolution of binding sites: assuming the sequence distribution $Q$ defines a motif at evolutionary equilibrium, its sequence information $H$  is proportional to the average fitness effect of its binding sites, $H (Q | P_0) = N \langle F \rangle$, with a proportionality constant equal to the effective population size~\cite{MCKLE,BL,MCPIE,BWL}. The fitness contribution of a particular binding sequence, $F(\a)$, is proportional to its log-likelihood ratio in the distributions $Q$ and $P_0$. The ensemble of these fitness values defines an information-based {\em fitness landscape} $F$ for binding of a specific transcription factor. These relations between sequence statistics and fitness of binding sites quantify our intuition that  specific sequences are overrepresented in a motif, because they confer a selective advantage over random sequences~\cite{SF}. If we write the motif distribution $Q$ in the product form of a position weight matrix, we obtain an approximate expression for the fitness $F(\a)$ in terms of the position-specific single-nucleotide frequencies $q_i (a)$ in the motif sequence and their counterparts $p_0(a)$ in background sequence: $N F(\a) =  \sum_{i=1}^\ell \log[q_i (a_i) / p_0 (a_i)]$. This expression, which is in its simplest form already contained in Kimura's U-shaped equilibrium distribution for a two-allele locus~\cite{Kimura}, is known as Bruno-Halpern model in the context of protein evolution~\cite{HB} and has been used to infer fitness effects of mutations in binding sites~\cite{MCKLE,BL,MCPIE,BWL,ML09,LASSIG}. Although this additive fitness model neglects fitness interactions between nucleotides within binding sites as well as between sites within a regulatory module, it is justified for the purpose of this study (see below). 

The fitness landscape $F$ defines the selection coefficient of any change from a state $\a$ to a state $\b$ of a binding site, $\Delta F_{\a \b} = F(\b) - F(\a)$. Together with the effective population size and the mutation rates, these selection coefficients determine the evolutionary dynamics of binding sites. In particular,  the probability $G^\tau (\b | \a)$ of evolving from an ancestor site $\a$ to a descendent site $\b$ through a series of point substitutions within an evolutionary distance $\tau$ can be evaluated in an analytical way from the underlying substitution matrix~\cite{DEKM,ML05} (see Materials and Methods). 

\begin{figure}[!h]

\includegraphics[scale=1.3]{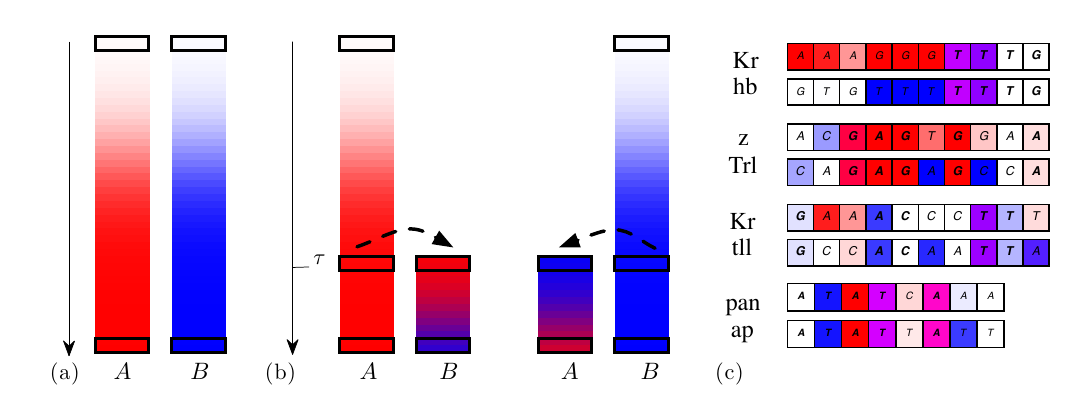}

\par
\caption{\label{fig:BSBSduplication} {\small
{\bf Evolutionary modes of transcription factor binding sites.} The figure shows alternative formation histories for two adjacent binding sites, whose present sequences bind transcription factors $A$ and $B$, respectively. The color coding indicates the evolution of binding function for factor $A$ (red) and $B$ (blue) with evolutionary time $t$. 
(a)~Evolution from independent ancestor sequences. The sites evolve to their present states by independent evolutionary processes under stationary selection given by different fitness landscapes $F_A$ and $F_B$ (see text). In this mode, adjacent sites will show no enhanced average sequence similarity compared to the similarity of their motifs.
(b)~Evolution by duplication of a common ancestor sequence. Left panel: The original site evolves in the stationary fitness landscape $F_A$. At a distance $\tau$ from the present, this site undergoes a duplication. The duplicated site evolves its new function of binding $B$ in the fitness landscape $F_B$. Right panel: The same process with the roles of $A$ and $B$ interchanged. In the duplication mode, the sites retain an enhanced sequence similarity, which reflects their common descent. 
(c)~Examples of adjacent functional binding sites with enhanced sequence similarity in the {\em D.~melanogaster genome}. The sites of each pair are aligned.  The color background of nucleotide $a$ at position $i$ indicates its contributions to fitness (binding affinity) for factor $A$ and $B$, i.e.,  $f_{i,A}(a)$  (level of red) and $f_{i,B} (a)$ (level of blue). The sequence similarity leads to hybrid binding characteristics: some nucleotides of the $A$-site (top row) have binding characteristics of the $B$-motif, and vice versa.  Examples from top to bottom (factor $A$ / factor $B$, genomic positions, duplication score):
(i)~{\em Kruppel} / {\em hunchback}, 
 chr3L: 8639822 / 8639878, 
$S = 4.40$,  
(ii)~{\em zeste} / {\em Trithorax-like},  
 chr3R:  12560236 / 12560218, 
$S = 3.97$, 
(iii)~{\em Kruppel} / {\em tailless}, 
 chr3L: 8639586 /  8639596, 
$S = 3.40$,
(iv)~{\em pangolin } / {\em apterous}, 
 chr3R: 22997722 / 22997752, 
$S=2.38$.}}
\end{figure}

We can now compare the two modes of binding site evolution introduced above. For any given pair of adjacent sites $\a$ and $\b$ that bind transcription factors $A$ and $B$, respectively, we want to evaluate the likelihood of two different histories of site formation. In the first mode of evolution, the sites are assumed to evolve to their present sequence states by point substitutions from independent ancestor sequences and under independent selection given by the fitness landscapes $F_A$ and $F_B$, as illustrated in Fig.~\ref{fig:BSBSduplication}(a). If the selection for binding is assumed to act over a sufficiently long evolutionary time, the probability of observing the present sequence states $\a$ and $\b$ in this independent mode of evolution is simply $Q_A(\a) Q_B (\b)$. This mode of evolution can only result in distance-dependent sequence similarity arising from an increased coverage with pairs of adjacent sites with correlated motifs $Q_A$ and $Q_B$ (evidence for this effect will be discussed below). However, it does not generate increased similarity of individual pairs of adjacent sites beyond that of their motifs.

In the second mode of evolution, the sites are assumed to evolve from a common ancestor sequence by a local duplication event at a distance $\tau$ from the present, followed by diversification under selection given by separate fitness landscapes $F_A$ and $F_B$: either the original site is under stationary selection for binding factor $A$ and the duplicated site has evolved the new function of binding the $B-$factor or vice versa, as illustrated in Fig.~\ref{fig:BSBSduplication}(b). In this mode, the present sequences $\a$ and $\b$ have evolved from their last common ancestor $\c$ by independent substitution processes with transition probabilities $G^\tau_A (\a | \c)$ and $G^\tau_B (\b | \c)$. The dynamics results in a joint probability of the form $
\small Q^{\tau}(\mathbf{a},\mathbf{b})=
\sum_{\mathbf{c}} G^{\tau}_{A}(\mathbf{a}|\mathbf{c})G^{\tau}_{B}(\mathbf{b}|\mathbf{c})Q(\mathbf{c}),
$
where the distribution of the ancestor sequence is given by $Q(\c) = [Q_A(\c) + Q_B(\c)]/2$ (see Materials and Methods). In this mode, distance-dependent sequence similarity arises due to common descent, causing the sequences of adjacent sites to be more similar than their motifs $Q_A$ and $Q_B$. Importantly, this effect is generic and not tied to any functional properties of the transcription factors $A$ and $B$. Fig.~\ref{fig:BSBSduplication}(c) shows a few examples of enhanced sequence similarity in pairs of adjacent binding sites in regulatory modules of {\em D.~melanogaster}.

The relative likelihood of common versus independent descent for a specific pair of sites $\a, \b$ is given by the {\em duplication score} $ S (\a, \b) = \log [Q^{\tau} (\a, \b) / Q(\a) Q(\b)]$. This score measures the similarity in a gapless alignment between the sequences $\a$ and $\b$ in a specific way: it gauges matches and mismatches depending on the weights of aligned nucleotides in their respective  binding motifs $Q_A$ and $Q_B$. The duplication score depends on the parameter $\tau$, which we choose by a maximum-likelihood procedure (see Materials and Methods). This parameter describes the expected excess similarity of site pairs related by common descent,  but it is not a linear clock of divergence time. Simulated evolution of binding site histories shows that the maximum-likelihood duplication score reliably distinguishes between site pairs $(\a,\b)$ with common and with independent descent (see Materials and Methods). Below, we use the distribution $W(S)$ of  duplication scores to infer the mode of evolution prevalent in a given class of site pairs.

This likelihood analysis goes beyond the inference of the sequence similarity $\K_\ell (r)$ introduced above. It can be seen as a decomposition of the distance-dependent similarity between sites into two parts:  the similarity between their motifs, and the excess similarity of the actual site pairs beyond that of their motifs. The first part reflects functional correlations within regulatory modules and is assigned to the background model $Q(\a) Q(\b)$.  Only the second part provides evidence for common descent, which is gauged by the scoring function $S(\a,\b)$. 

Our model scores only the sequence similarity within site pairs and does not incorporate the insertions and deletions between the sites after duplication, which determine their relative distance. 
This is justified, because the likely divergence times of most duplicated site pairs are much longer than repeat lifetimes. If a site duplicates within a repeat, the relative distance between copies may subsequently undergo rapid evolution due to the high indel rates in these regions [46--49]. Given a surplus of insertions over deletions in regulatory modules,  we expect the relative distance to increase on average [48]. The spacing of contemporary sites is then the result of a long-term diffusive insertion/deletion dynamics within the repeats active since duplication, most of which have decayed in today's sequence. This leaves the similarity of conserved functional sites as the most prominent long-term marker of these dynamics.

\subsection*{Local sequence duplications in {\em Drosophila}}
\begin{figure}[T]
\includegraphics[scale=0.32]{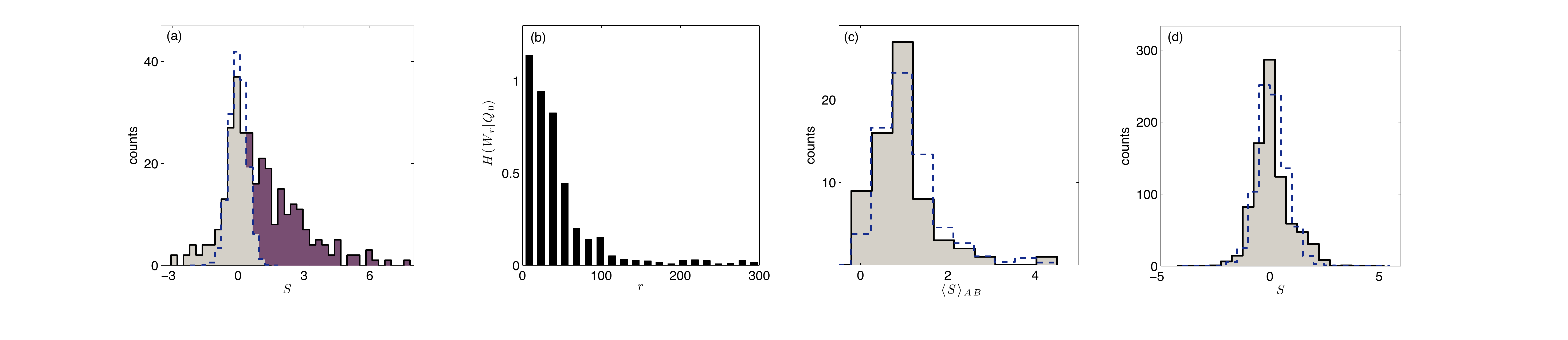}
\par
\caption{\label{fig:scoredist}{\small
{\bf Common vs. independent descent of binding sites in fly and yeast.} 
(a)~Histogram of the duplication score $S$ for $306$ pairs of binding sites with a mutual distance of up to $50$ bp in the genome of {\em D.~melanogaster} (sum of grey-shaded and violet-shaded part).  Decomposition of counts according to the mixed-descent model (see Materials and Methods): 43\% of the site pairs are of independent descent and have the score distribution $Q_0(S)$ (obtained from pairs with relative distance $r>200$ bp, dashed line), 57\% of the site pairs of are of common descent and have the score distribution $Q(S)$ (violet-shaded). 
(b)~Relative log-likelihood for duplication per site pair, i.e., relative entropy $H(W_r|Q_0)$ obtained from the score distribution $W_r(S)$ of site pairs in the relative distance range $(r, r + 15)$ bp (evaluated from a total of $506$ sites). The rapid decay of this function suggests a local mechanism generating excess similarity between adjacent sites. 
(c)~Histogram of partial score averages $\langle S \rangle_{AB}$ for all factor pairs $(A,B)$ binding the site pairs of~(a) (grey-shaded) and corresponding distribution of averages obtained after scrambling the score values of site pairs (normalized to the same number of total counts, dashed line). The two distributions are statistically indistinguishable (KS-test {\em p-}value = 0.8378), which shows that positive duplication scores are not limited to a subset of factor pairs. 
(d)~Histogram of the duplication score $S$ for $833$ pairs of binding sites with a mutual distance of up to 50 bp in the genome of {\em S.~cerevisiae} (grey-shaded). The distribution is not significantly different from the null distribution obtained from random site pairs (normalized to the same number of total counts, dashed line), i.e., there is no evidence for common descent as prevalent evolutionary mode. }}
\end{figure}

Using the duplication score $S$, we have evaluated the sequence similarity of $506$ pairs of neighboring binding sites in regulatory modules of the \textit{Drosophila melanogaster} genome. These sites are experimentally validated and recorded in the REDfly database~\cite{HGB,GLHH,BCC} (see Materials and Methods).  We infer the prevalent mode of evolution as a function of the distance $r$ between sites and obtain the main result of this paper:

\noindent
---{\em In fly, binding sites with a distance of up to about $50$ bp are more likely to share a common ancestor than to have evolved from independent origins.}  Fig.~\ref{fig:scoredist}(a) shows the histogram of duplication scores $S (\a,\b)$ for the set of $k = 306$ binding site pairs with $r \leq 50$ bp. The score distribution $W(S)$ of these pairs is clearly distinguished from the background distribution $Q_0(S)$, which is obtained from pairs of sites located in the same module at a distance $r > 200$ bp and is associated with independent descent.  We decompose the score distribution of adjacent sites in the form $W(S) = (1 - \lambda) Q_0(S) + \lambda Q(S)$, attributing the excess of large scores  to pairs of sites of common descent with a score distribution $Q(S)$. Our best fit of this mixed-descent model to the data distribution has a fraction $\lambda = 57 \%$ of adjacent site pairs formed by duplication; see Fig.~\ref{fig:scoredist}(a). The total log-likelihood of the mixed-descent model relative to the background model is given by multiplying the relative  entropy of the distributions $W$ and $Q_0$ with the number of site pairs, $\Sigma = k \, H (W | Q_0)$. We estimate $\Sigma > 234$, providing significant statistical evidence that the prevalent mode in adjacent sites is evolution from common descent (for details, see Materials and Methods). We note that this significance emerges for the ensemble of the adjacent site pairs, whereas the relative log-likelihood for duplication per site pair, $H(W|Q_0)$, is of order one: individual site sequences are inevitably too short to reliably discriminate between the two evolutionary modes.
Our conclusion that local sequence duplications generate the observed excess similarity of adjacent sites is supported by a number of further controls and a comparison with the yeast intergenic regulatory sequences:

\noindent
---{\em The relative log-likelihood for duplication per site pair decreases with increasing distance $r$ between sites.} In Fig.~\ref{fig:scoredist}(b), we evaluate the relative entropy $H(W_r | Q_0)$ for the score distributions $W_r (S)$ of site pairs with different values of mutual distance $r$. We find a rapid decay up to about $100$ bp, that is, the score distribution $W_r$ becomes successively more similar to the background distribution $Q_0$ with increasing site distance. This pronounced distance-dependance is comparable to that of  the total sequence similarity shown in Fig.~\ref{fig:Infoauto}(c) and is consistent with local duplications as underlying mechanism.

\noindent
---{\em Similarity of neighboring sites is not limited to specific pairs of transcription factors.} We partition the 306 site pairs with a mutual distance of less than 50 bp by factor pairs and evaluate the partial score averages $\langle S \rangle_{AB}$. We compare the  distribution of these averages with the corresponding distribution of averages evaluated after scrambling the score values of the site pairs, as shown in Fig.~\ref{fig:scoredist}(c). The two distributions are statistically indistinguishable, which shows that excess sequence similarity is a broad feature of adjacent binding sites and is not limited to a subset of sites for factor pairs with specific functional relationships. This supports our conclusion that the excess sequence similarity reflects common descent and not fitness interactions (epistasis) between sites. Of course, epistasis is common for binding sites in the same regulatory module, because these sites perform a common regulatory function. However, generic interactions couple the binding energies of adjacent sites, not directly their sequences. Epistatic effects generating excess sequence similarity are conceivable for specific factor pairs, but do not appear to be a parsimonious explanation for the broad similarities of adjacent binding sites we observe. 

\noindent
---{\em In yeast, binding site duplications are not frequent.}   
 For comparison, we have also evaluated a set of $1352$ pairs of binding sites in the {\em Saccharomyces cerevisiae} genome. Fig.~\ref{fig:scoredist}(d) shows distribution of duplication scores $S(\a,\b)$ for the set of binding sites with $r \leq 50$ bp. This distribution is strongly peaked around zero (because the maximum-likelihood value of $\tau$ is large, see Materials and Methods) and indistinguishable from the distribution of the control set of random site pairs; both distributions have a negative average. As in {\em Drosophila}, most binding sites in the same intergenic region of {\em S. cerevisiae} are located within $50$ bp from each other. However, we do not observe evidence for local duplications as a mode of binding site formation in yeast. Clearly, this result does not exclude that such duplications take place, but they do not appear to be frequent enough to generate a statistically significant excess similarity of neighboring sites. This is not surprising given the differences in regulatory architecture between yeast and fly: individual sites in {\em S. cerevisiae}  are more specific than in {\em Drosophila};  the average sequence information of a binding motif is  $H \approx 12-17$ bits, compared to $H\approx 6-8$ bits~\cite{WM}. Accordingly, a larger part of the regulatory functions in yeast relies on single sites, and there are no regulatory modules which would require frequent duplications for their formation. 

\subsection*{Adaptive potential of duplications}

Do the inferred site duplications have adaptive potential for the formation of novel binding sites? 
Here, we use the term adaptive potential to indicate that the duplication itself may be a neutral process, and selection for factor binding may latch on later to duplicated sites. The duplication of a site for a given transcription factor has obvious adaptive potential towards formation of an adjacent site for the same factor. But local duplications also have adaptive potential if the duplicated site is to evolve the new function of binding a different factor, because the binding motifs of transcription factors with adjacent sites are correlated. This correlation quantifies the ability of one factor to recognize the binding sites of another factor, including seed sites generated by sequence duplications. Specifically, we define the binding correlation $H_{AB}$ of a transcription factor $A$ with another factor $B$ as the average information-based fitness to bind factor $B$ in the ensemble of $A$-sites. In Fig.~\ref{fig:adaptiveadvantage}, this quantity is evaluated for all factor pairs $(A,B)$ with adjacent binding sites, together with the range of fitness $F_B$ of known target sites for factor $B$ and the average $F_B$ in background sequence (see Materials and Methods). For most such factor pairs, the fitness of a typical $A$-site is seen to be similar to that of weak $B$-sites and significantly larger than the average fitness of background sequence. This binding correlation between motifs is sufficient so that an $A$-site duplicate can act as a seed for a $B$-site, which can subsequently adapt its strength by point mutations. The binding correlation is specific to factors which have adjacent binding sites; we have found no such effect in the control ensemble of all factor pairs $(A,B)$ (most of which do not have adjacent sites). Furthermore, some  highly specific motifs, such as {\em hunchback}, {\em twi} and {\em z}  do not show binding correlations with other factors. 

\section*{Discussion}

\subsection*{Local sequence duplication as a mechanism of regulatory evolution}
Local sequence duplications (and deletions) are a generic evolutionary characteristic of intergenic DNA and, in particular, of regulatory sequence~\cite{SS,BRPM,TS,VLCHV,MA,Gruen}.
 In this study, we have established evidence for local sequence duplications as a mechanism that transports and produces cis-regulatory information. These duplications generate specific, distance-dependent sequence similarity in strongly correlated pairs of sites with a relative distance of up to about 50 bp, which account for a substantial part of the sequence autocorrelation in fly regulatory modules. In particular, they provide a parsimonious explanation for the observed excess sequence similarity of  transcription factor binding sites in this range of relative distance. We conclude that the majority of these adjacent site pairs have a common ancestor sequence. The large amplitude of the duplication signal may be the most surprising result of this study. It far exceeds the level expected from the repeats in contemporary sequence, which cover only about 5 percent of binding sites and are typically shorter than the distance between correlated sites. Common-descent site pairs are the cumulative effect of past duplications over macro-evolutionary intervals, whose trace is conserved by selection on site functionality. 
 
 This result establishes local duplication as a pervasive formation mode of regulatory sequence, which generates, for example, the known local variations in site numbers between {\em Drosophila} species. Of course, our evidence for this mode is statistical and, at this point, is confined to a limited dataset of binding sites with confirmed functionality~\cite{HGB,GLHH,BCC}.
  The duplication mode appears to be specific to multicellular eukaryotes; we have not found comparable evidence in the yeast genome.  
Our findings are relevant for genome analysis in two ways: including local duplications should inform inference methods for binding sites as well as alignments of regulatory sequence with improved scoring of indels~\cite{TS,MA,SS,Gruen}. With such methods, it may become possible to follow the evolutionary history of binding sites in more detail.  

\subsection*{Life cycle of a binding site}
We have found evidence that local duplications can confer adaptive potential for the formation of novel binding sites, because they generate seed sequences with marginal binding specifically in the vicinity of existing sites. This mechanism is necessary, because point mutations alone can only lead to rapid loss but not to gain of  new sites with positional specificity. Thus, duplications and point mutations complement each other, suggesting that typical binding sites within multicellular eukaryotes have an asymmetric life cycle: formation within a functional cluster by local duplication, adaptation of binding energy by point mutations, evolution of relative distance to neighboring sites by insertions and deletions in flanking sequence, conservation by stabilizing selection on binding energy, and loss by point mutations.

The life cycle of individual binding sites interacts with other levels of genome evolution. Gene duplications with subsequent sub-functionalization have been identified as an important evolutionary mode specifically in higher eukaryotes~\cite{LC}. If subfunctionalization is initialized at the level of gene regulation, it amounts to a loss of regulatory input for both gene duplicates and provides a mechanism for adaptive loss of binding sites. This process alone would lead to genomes with many genes, but few functions per gene. Maintaining regulatory complexity with multi-functional genes as observed in eukaryotic genomes ~\cite{Ptashne,DAVIDSON} requires a converse evolutionary mode: gain of new functions by existing genes. At the regulatory level, this amounts to gain of regulatory input, i.e., adaptive formation of new binding sites.

\subsection*{Sequence evolution and regulatory grammar}
Previous studies have identified regulatory modules as important units of transcriptional control, in which clusters of binding sites bind multiple transcription factors with cooperative interactions. The sites in a cluster follow a regulatory grammar resulting from natural selection acting on site order, strength, and relative distances~\cite{SAL,MMML,KA}.  If sequence duplications play a major role in the formation of such clusters, we may ask how much of their observed structure reflects this mode of sequence evolution, rather than optimization of regulatory function by natural selection. Local duplications generically produce descendant sites, which are weak binding sites for another factor at best, as shown in  Fig.~\ref{fig:adaptiveadvantage}. (Significant heterogeneity in binding strength between adjacent sites is indeed observed in our sample.) The resulting binding sequences are hardly optimal in terms of specificity and discrimination between different factors. Cooperative binding between transcription factors may have evolved as a secondary mechanism to confer regulatory function to these sequence structures. 

In this paper, we have argued that local sequence duplications facilitate the adaptive evolution of gene regulatory interactions. However, the adaptive potential of duplications does not imply that the duplication process itself has to be adaptive or even confined to regulatory sites. Similar to gene duplications~\cite{LC}, many site duplications may be neutral and provide a repertoire of marginal regulatory links. Adaptive diversification can build subsequently on this repertoire, conserving and tuning those links that confer a fitness advantage  and discarding others. 

\section*{Materials and Methods}

\subsection*{Regulatory sequences and position weight matrices}

The sequence analysis of \textit{D.~melanogaster} is based on the  \textit{cis-}regulatory modules and experimentally validated binding sites collected in the REDfly v$.2.2$ database \cite{HGB,GLHH,BCC}, and on the position weight matrices of Dan Pollard's dataset~({\sl http://www.danielpollard.com/matrices.html}).
To measure the distance-dependent sequence similarity $\K_\ell (r)$, we use the 346 known regulatory modules with length of more than 1000 bp in  \textit{D.~melanogaster}. The analysis in \textit{D.~yakuba} and \textit{D. pseudoobscura} is based on the 249 well-aligned homologous regions obtained from multiple alignments of  12  Drosophila species (dm3, BDGP release5); see Fig.~\ref{fig:crossSpec}. For the evolutionary inference in the second part of the paper, we use only the experimentally validated binding sites contained in these modules which are not necessarily selected for high similarity to motifs or for high mutual similarity.  To avoid biases in our analysis, the set of sites is truncated in three ways: (i)~We only use binding sites for transcription factors that occur in at least two different regulatory modules, so that the position weight matrix is not biased by the sequence context of a single module. (ii)~We use only sites that have no sequence overlap with other sites in the dataset, because our inferred fitness landscapes describe the selection for a single regulatory function~\cite{MKCL}. (iii)~We exclude sites in the {\em X} chromosome, which could bias the results by its high rate of recent gene duplications and the abundance of repeat sequences~\cite{KATTI,TL}. These conditions produce a cleaned set of 506 transcription factor binding site pairs located in 74  \textit{cis-}regulatory modules.
For the analysis in {\em S.~cerevisiae}, we use sites and position weight matrices from the SwissRegulon database~\cite{PEMv}. 
These footprints do not always match the length $\ell$ of their position weight matrices. To produce a set of site sequences of common length $\ell$, longer footprints are cut and shorter ones joined with flanking nucleotides, such that the binding affinity is maximized.

\subsection*{Sequence information measures}
Sequence autocorrelation is a measure of enhanced mean similarity between the nucleotides of a sequence segment. The distance dependence of the autocorrelation signal provides information about  the range, within which the nucleotides appearing in the sequence are correlated. In a given sequence segment $a_1, \dots, a_L$, the nucleotide frequencies are given by 
\EQ
p_0 (a) =\frac{1}{L} \sum_{\nu =1}^L \delta (a_\nu, a), 
\label{p0}
\EE
where $\delta(a_\nu,a) = 1$ if $a_\nu = a$ and $\delta(a_\nu,a) = 0$ otherwise. These determine the mean similarity between two random nucleotides of the segment, $c_0 = \sum_a p_0^2 (a)$.  The sequence autocorrelation function is then defined by 
\EQ
\Delta (r) =  - c_0 +\frac{1}{L - r} \sum_{\nu = 1}^{L-r} \delta (a_\nu, a_{\nu + r}).
\label{Delta}
\EE
We evaluate this function in the 346 regulatory modules of {\em Drosophila melanogaster} genome with length of more than 1000 bp identified by REDfly v$.2.2$ database \cite{HGB,GLHH,BCC}. As shown in Fig.~1(a), we find an approximate exponential decay with a characteristic length of about 100 bp as the range of sequence correlation. The mean identity $c_0$ is evaluated in a local window of 500 bp (changing the window length affects the baseline of this function, but not its dependence on distance up to 100 bp). Information about the spatial distribution of these correlated nucleotides along the genome is contained in higher orders of sequence autocorrelation (i.e., reoccurrence of doublets, triplets, etc.). 
Here, we use information theory to identify such clusters of correlated nucleotides in a sequence region.\\
We want to detect reoccurring nucleotide patterns or {\em motifs}. A motif of length $\ell$ is a probability distribution $Q(\a)$ for sites $\a = (a_1, \dots, a_\ell)$ which differs significantly from the background distribution $P_0(\a)$. If we neglect correlations between nucleotides, we can write these distributions as the product of single-nucleotide frequencies, 
\EQ
Q(\a) = \prod_{i=1}^\ell q_i (a_i)
\label{Q}
\EE
and $P_0(\a) = \prod_{i=1}^\ell p_0(a_i)$. The $4 \times \ell$ matrix of single-nucleotide frequencies (\ref{Q}) is called the position weight matrix of the motif. The {\em sequence information} of the motif is defined as the relative entropy (Kullback-Leibler distance) of these distributions \cite{KL}, 
\EQ
H(Q|P_0) = \sum_{i =1}^\ell \sum_a q_i (a) \log \frac{q_i(a)}{p_0(a)}. 
\label{H}
\EE
To study the  sequence coverage by informative motifs, we use a reduced form of the full frequency distribution $Q$ by mapping it  to the mean similarity of its motif sites. Hence, even without any prior knowledge on frequency distributions,  we can recover part of the sequence information for those motifs that occur more than once in the sequence segment. Two sites drawn from the motif have a mean similarity $c = \sum_{i,a} q_i^2(a)$ between aligned nucleotides, which is higher than the background mean similarity $c_0$. The {\em similarity information} of the motif is given by the relative entropy  
\EQ
K (c, \ell |c_0) = \ell \left [c \log \frac{c}{c_0} + (1 - c) \log \frac{1 - c}{1 - c_0} \right ]. 
\label{K}
\EE
Similarity information between pairs of sites is a somewhat diluted measure  of sequence information. As a rule of thumb, the mutual similarity entropy per site pair, $K(c, \ell | c_0)$, recovers about half of the motif information per site,  $H(Q| P_0)$. 

\subsection*{Inference of similarity information by dynamic programming}

To estimate the total similarity information $K_\ell (r)$ of all strongly correlated pairs of sites with distance $r$ and length $\ell$ in a sequence segment of length $L$, we construct a set 
\EQ
\{(a_{\nu_1}, \dots, a_{\nu_1 + \ell -1}), (a_{\nu_1 + r}, \dots, a_{\nu_1 + r + \ell - 1})\}, \dots,
\{(a_{\nu_n}, \dots, a_{\nu_n + \ell -1}), (a_{\nu_n + r}, \dots, a_{\nu_n + r + \ell - 1})\}
\label{pairset}
\EE
of site pairs with the following properties: \\
(i) The left (and also the right) sites of all pairs have no sequence overlap, 
\EQ
\nu_{\alpha + 1} - \nu_\alpha \geq \ell \;\;\; \mbox{ for $\alpha = 1, \dots, n - 1$}.
\EE
(ii) The mean similarity of each pair is greater than a threshold $c_{\min}$,
\EQ
c_\alpha \equiv \frac{1}{\ell} \sum_{i =1}^\ell \delta (a_{\nu_\alpha + i}, a_{\nu_\alpha + r + i}) >  c_{\min}
\;\;\; \mbox{ for $\alpha = 1, \dots, n$}.
\EE
(iii) The sum of mutual similarities $\sum_{\alpha = 1}^n c_\alpha$ is maximal (see Fig.~\ref{fig:frameshift}) 

\noindent By the dynamic programming recursion 
\EQ
C_t = \max[C_{t-1},  C_{t-\ell} + [\frac{1}{\ell} \sum_{i=1}^\ell \delta(a_{t-\ell + i -r}, a_{t-\ell + i})]-c_{\min}], 
\EE
we obtain the sequence of partial scores $C_1, \dots, C_L$ with the initial condition $C_1 = 0$.  We then use a backtracking procedure~(see, e.g., \cite{DEKM}) to determine the set of positions $(\nu_1, \dots, \nu_n)$ and, hence, the number $n(r,\ell, c_{\min}) $ and the average similarity $\bar c (r,\ell, c_{\min}) = (C_L/n) + c_{\min} $  of the high-similarity pairs (\ref{pairset}). To estimate  the expected number of pairs in background sequence,  $n_0(r,\ell,c_{\min})$,  we apply the same procedure to $1000$ sequences of length $L$, which are generated by a first-order Markov model
\EQ
P(a_1, \dots, a_L) = p_0 (a_1) \prod_{\nu = 2}^L T(a_\nu | a_{\nu -1})
\EE
with the same single-nucleotide frequencies $p_0(a)$ and conditional frequencies $T(a|b)$ as in the actual sequence. We then evaluate the excess $\Delta n (r,\ell, c_{\min}) = n (r,\ell, c_{\min}) - n_0 (r,\ell, c_{\min})$ and obtain an estimate of the total information contained in the enhanced autocorrelation of motifs as given by eq.~(\ref{K}),
\EQ
\K_{\ell} (r) = \ell \max_{c_{\min}}  \left [\frac{ \Delta n (r,\ell, c_{\min})}{L}  \left ( \log \frac{\bar c (r,\ell, c_{\min}) }{c_0} + \log \frac{1 - \bar c (r,\ell, c_{\min})}{1 - \bar c_0}\right ) \right ]. 
\EE
We  infer $c_{\min}$ by maximum likelihood analysis of the total similarity information in the sequence. This method also allows for optimization of the motif length $\ell$,  similar to the procedure in the local sequence alignment algorithms~\cite{DEKM}.

\subsection*{Evolutionary model for binding sites}

Our evolutionary dynamics of binding site sequences $\a = (a_1, \dots, a_\ell)$ for a given transcription factor is determined by the Bruno-Halpern fitness model~\cite{HB} derived from the position weight matrix $q_i (a)$ ($i = 1, \dots, \ell$)  and the background frequencies $p_0(a)$, 
\EQ
N F(\a) = \sum_{i=1}^\ell f_i (a_i) \quad \text{with}\ f_i(a) = \log \frac{q_i(a)}{p_0(a)}.
\label{FBH}
\EE
This relationship between fitness and nucleotide frequencies is valid if binding sites are at evolutionary equilibrium under mutations, genetic drift, and selection, and background sequence is at neutral equilibrium (accordingly, all inferred fitness values are scaled in units of the effective population size $N$). The relationship of the evolutionary ensembles with the underlying thermodynamics of site-factor interactions is discussed, for example, in ref.~\cite{LASSIG}. Eq.~\ref{FBH} defines an information-based fitness model: the average fitness of functional binding sites equals the sequence information of the motif, 
\EQ
\langle F \rangle = H(Q | P_0)
\EE
with $Q(\a) = \prod_{i=1}^\ell q_i(a_i)$ and $P_0(\a) = \prod_{i=1}^\ell p_0(a_i)$; see eqs.~(\ref{p0}), (\ref{Q}) and (\ref{H}). We infer  $p_0(a)$ from the local background frequency of the region $500$ base pairs around each binding site.  The rates $u_{\a \to \b}$ of point substitutions $\a \to \b$ within binding sites are determined by the scaled selection coefficients $N \Delta F_{\a \b} = N[ F(\b) - F(\a)]$ derived from this fitness model and the point mutation rates $\mu_{\a \to \b}$ (which are assigned a uniform value $\mu$ for simplicity). Here, we use the standard Kimura-Ohta substitution rates
\EQ
u_{\a \to \b} =  \mu_{\a \to \b} \frac{N \Delta F_{\a \b}}{1 - \exp (- N \Delta F_{\a \b})},
\label{u}
\EE
which are valid in the regime $\mu N \ll 1$ (in which subsequent substitution processes are unlikely to overlap in time) and $\Delta F_{\a\b} \ll 1$~\cite{Kimura,KO}. The matrix of these substitution rates  
then determines the transition probabilities (propagators) $G^\tau (\b | \a)$ from an arbitrary initial  sequence  $\a$ to an arbitrary final sequence $\b$ within an  evolutionary distance $\tau$~\cite{DEKM,ML05}.  Given the set of transition probabilities, we obtain the joint probability $Q^\tau (\a,\b)$ for a pair of sites $(\a, \b)$ that bind transcription factors $A$ and $B$, respectively, and have evolved from a common ancestor $\c$ as described in the main text and in Fig.~2(b). First, we assume that the ancestor site is at evolutionary equilibrium under selection to bind factor $A$, that is, the contemporary site $\a$ has the ancestral function and $\b$ has evolved a new function after duplication. This gives the contribution
\begin{eqnarray}
Q_A^\tau (\a, \b) & = & \sum_\c G^\tau_A(\a |\c) G^\tau_B (\b | \c) Q_A(\c) 
\nonumber \\
& = & \sum_\c G^\tau_B(\b |\c) G^\tau_A (\c | \a) Q_A(\a), 
\label{Qa;b}
\end{eqnarray}
where we have used the detailed balance condition of the substitution dynamics~\cite{ML05}. There is a second contribution $Q_B^\tau (\b,\a)$ describing the case of the ancestor $\c$ under stationary selection to bind factor $B$. Weighing these cases with equal prior probabilities, we obtain 
\EQ
Q^\tau (\a, \b) = \frac{1}{2} \big [ Q_A^\tau (\a,\b) + Q_B^\tau (\b, \a) \big ].
\label{Qtau}
\EE
In our analysis of pairs of adjacent binding sites in {\em Drosophila}, there is usually a  dominant contribution from one of the terms, from which we can infer the likely function of the ancestor site. In the limit of large $\tau$, the evolution from a common ancestor becomes indistinguishable from evolution by independent descent, $\lim_{\tau \to \infty} Q^\tau (\a, \b) = Q_A(\a) Q_B(\b)$. 

\subsection*{Inference of common descent}

The duplication score
\EQ
S(\a, \b)=\log \frac{Q^\tau (\a, \b)}{ Q_A(\a) Q_B(\b)}
\label{Sdupl}
\EE
is a measure of sequence similarity between binding sites. This score depends on the evolutionary distance parameter $\tau$. We infer the optimal value of $\tau$ by maximizing the likelihood ratio between the score distribution of pairs with mutual distance $r<50$ and the score distribution of pairs with independent origin. In {\em D. melanogaster}, we find a finite maximum-likelihood evolutionary distance $\tau \approx 0.4 \mu^{-1}$ and significantly positive values of the duplication score for adjacent binding  sites. In {\em S.~cerevisiae}, we find large values $\tau \gg \mu^{-1}$, i.e., there is no statistical evidence for evolution by common descent.  Our conclusions are largely independent of the values of $\tau$ used in (\ref{Qtau}) and (\ref{Sdupl}). These values should be regarded as model fit parameters for the observed sequence similarities. Energy-based fitness models~\cite{ML05, MKCL}, which take into account the epistasis between mutations within binding sites, are required to obtain more accurate estimates of $\tau$, which can be tested against phylogenetic data. Epistasis will increase the inferred values of $\tau$ compared to the additive (Bruno-Halpern) model~\cite{ML05, MKCL}.\\

We evaluate the score distribution $W(S)$ of a given class of site pairs in terms of a mixture model of common and independent descent, 
\EQ
W (S) = (1 - \lambda) Q_0 (S) + \lambda Q(S).
\label{W}
\EE
The distribution of scores for independent descent, $Q_0(S)$, is obtained from pairs of sites in a common module with a relative distance $r > 200$ bp (Fig.~ 3(a), dashed line). This distribution is approximately Gaussian and has a width of order one, which is consistent with the simulations reported below. Because we build $Q_0$ from sites in a common module, its score average is above that for pairs of sites located in different modules. In this way, the overall sequence similarity within modules, which depends on the local GC-content, is  assigned to the background model and does not confound the evidence for common descent. The distribution $Q(S)$ is the best fit to the  the large-score excess of the distribution $W(S)$ for adjacent sites with a relative distance $r < 50$ bp (Fig. 3(a), violet-shaded). This distribution has larger mean and is broader than the background distribution $Q_0$, which is also consistent with the simulations reported below. 

Given a set of $k$ site pairs $(\a, \b)$ with scores $S(\a,\b)$ described by the distribution $W(S)$,  the log-likelihood of the mixed-descent model (\ref{W}) relative to the independent-descent background model is given by 
\EQ
\Sigma = k \, H(W | Q_0) 
= \sum_{\mbox{\small site pairs}} \log \left [ \frac{W(S(\a,\b))}{Q_0(S(\a,\b))} \right ]
= \sum_{\mbox{\small site pairs}} \log \left [ (1 - \lambda) + \lambda \frac{Q(S(\a,\b))}{Q_0(S(\a,\b))}  \right ]; 
\label{W2}
\EE
it equals the product of the number of sites and the relative entropy $H(W|Q_0)$. The extensive quantity $\Sigma$ measures the statistical evidence for the mixture model based on the number and the score distribution of site pairs, whereas $H(W|Q_0)$ quantifies only the shape differences between the distributions $W(S)$ and $Q_0(S)$. We evaluate eq.~(\ref{W2}) using the conservative estimate $Q(S)/Q_0(S) \geq \exp(S - S_0)$ with $S_0 = 0.7$; see Fig.~3(a).

We have tested our inference procedure by simulations of the sequence evolution for pairs of binding sites with common and with independent descent. For these simulations, we use four pairs of different factors $\{ A , B\} = \{ ftz ,bcd\}, \{ ftz ,abd-A\}, \{  bcd, abd-A\} , \{ bcd,  Kr\}$,  and two pairs of equal factors $\{ A , B\} = \{  ftz , ftz\}, \{  bcd, bcd\}$. For each factor pair, we obtain an ensemble of 25000 pairs of binding sites $(\a,\b)$ with a duplication in their evolutionary histories, as described by Eqs.~(\ref{Qa;b}, \ref{Qtau}) and Fig.~\ref{fig:BSBSduplication}(b). We first obtain 500 duplication events $(\c,\tau)$: the last common ancestor sequence $\c$ is drawn with equal likelihood from the ensemble $Q_A (c)$ or $Q_B(c)$, and the divergence time $\tau$ is drawn from an exponential distribution with mean $\bar \tau = 0.4/\mu$. For each duplication event, we draw 50 site pairs  $(\a,\b)$ from the distribution $Q^\tau_A (\a|\c) Q^\tau_B (\b | \c)$ describing evolution under selection for binding of factors $A$ and $B$, respectively. 
We then apply our scoring procedure to this set of site pairs. As for the real sequence data, we infer a single maximum-likelihood parameter $\tau_{\rm ML}$ by maximization of the total duplication score $\Sigma$. As shown in Fig.~\ref{fig:simulatedBS}(a), $\Sigma$ has a pronounced maximum at a value  $\tau_{\rm ML} \approx 0.3/\mu$, which is close to the mean divergence time $\bar \tau$ of the input data. 
We conclude that the constraint of a fixed $\tau$ does not confound the inference of common descent. We also obtain separate score distributions  for sites $(\a,\b)$ binding the pairs $(A,B)$ of equal factors and of different factors listed above; see Fig.~\ref{fig:simulatedBS}(b) and Fig.~\ref{fig:simulatedBS}(c). These distributions are similar and clearly distinguish duplicated site pairs from pairs with independent ancestries for both factor groups. We conclude that our method can infer common descent of binding sites, independently of their functional characteristics.

\subsection*{Binding correlation of transcription factors }
\begin{figure}[T]
\begin{center}
\includegraphics[scale=0.54]{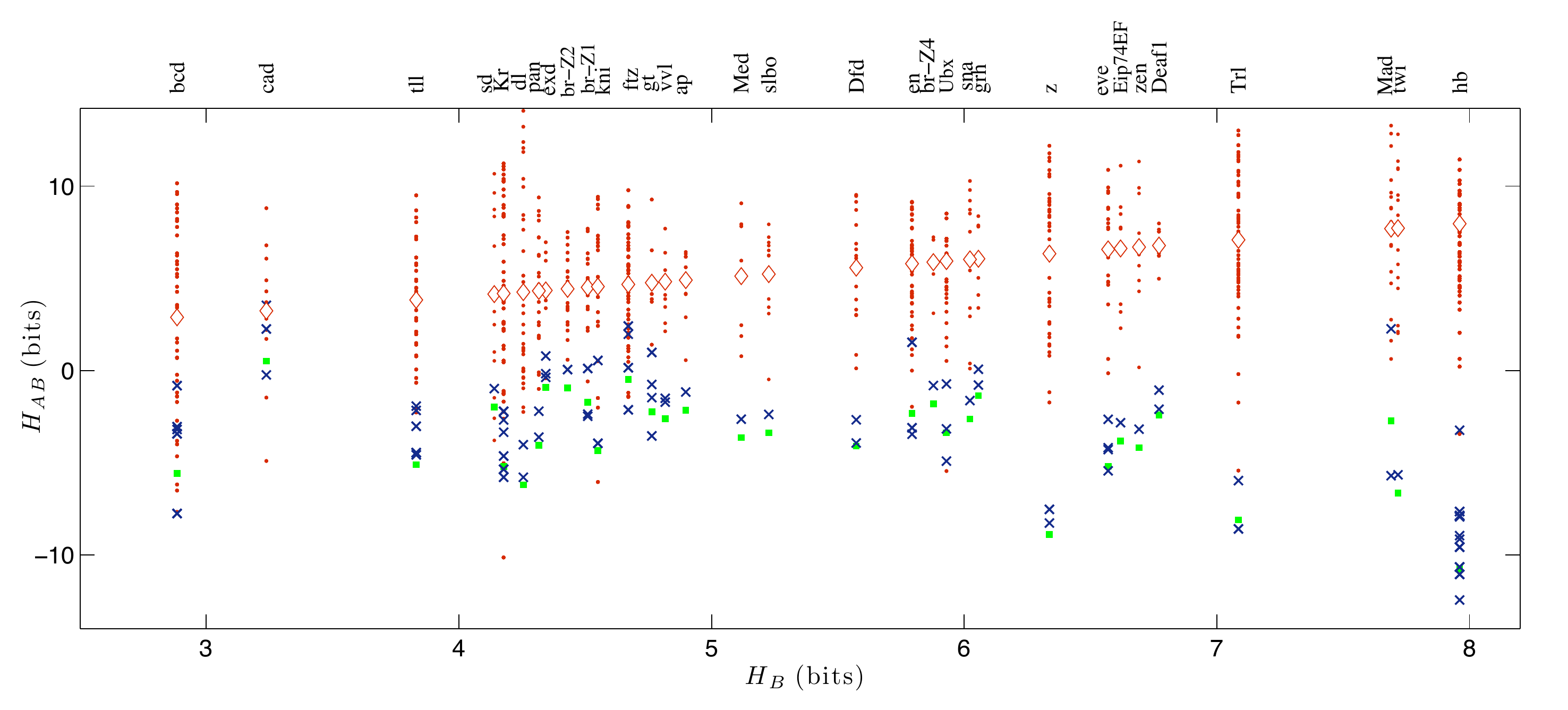}
\end{center}
\par 
\caption{\label{fig:adaptiveadvantage} {\small
{\bf Adaptive potential of binding site duplications.} 
The binding correlation $H_{AB}$ of all pairs of {\em Drosophila} transcription factors $(A,B)$ which have adjacent binding sites in a common regulatory module is evaluated as the average information-based fitness of $A$-sites for factor $B$ and plotted against the sequence information $H_B$ of the binding motif of factor $B$ (blue crosses); see eqs.~(\ref{HAB}) and (\ref{HB}) in Materials and Methods. The binding correlation is compared to the distribution of fitness values $F_B$ of the $B$-sites (red dots, the average fitness for each factor is shown as diamond and equals the abscissa $H_B$) and to the average fitness $F_B$ in background sequence (green dots); see  eq.~(\ref{H0B}) in Materials and Methods. The binding correlation $H_{AB}$ is significantly larger than the background average of $F_B$ and is comparable to the fitness $F_B$ of weak $B$-sites in a substantial fraction of cases.  Some  highly specific motifs, such as {\em hunchback}, {\em twi} and {\em z}  do not show binding correlations with other factors.
}}
\end{figure}

We define the binding correlation $H_{AB}$ for each ordered pair of factors $(A,B)$ as the average information-based fitness of $A$-sites for the $B$-factor,
\EQ
H_{AB} = \langle F_B \rangle _A = \sum_{i,a}q_{A,i}(a)f_{B,i}(a)  \quad \text{with}\quad f_{B,i}(a)=\log \frac{q_{B,i}(a)}{p_0(a)}.
\label{HAB}
\EE
This value is an estimate for the compatibility of the $A$-sites with the transcription factor $B$ and equals, up to a constant, the information-theoretic {\em cross entropy} between the distributions $Q_A$ and $Q_B$. In Fig.~\ref{fig:adaptiveadvantage}, this quantity is compared to (i) the sequence information $H_B$ of the motif $Q_B$, which equals the average fitness of $B$-sites for the $B$-factor by eq.~(\ref{H}),
\EQ
H_B \equiv H(Q_B |P_0) = \sum_{i,a}q_{B,i}(a)f_{B,i}(a), 
\label{HB}
\EE
and (ii) to the average fitness of background sequence for the $B$-factor, 
\EQ
H_{0B} = \sum_{i,b}p_0(b) f_{B,i} (b).
\label{H0B}
\EE

\newpage{}

\setcounter {figure}{0}

\def\thefigure{S\arabic{figure}}
\begin{figure}[T]
\raggedright{}\includegraphics[scale=0.47]{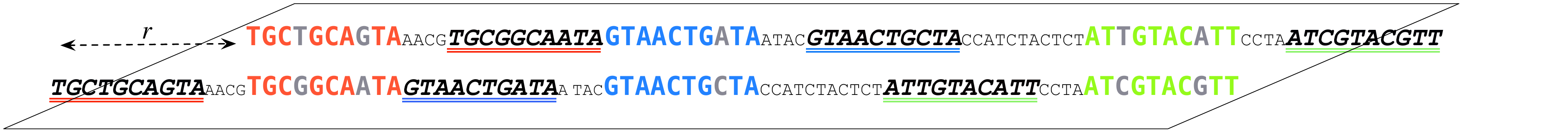}
\par
\caption{\label{fig:frameshift} {\small {\bf Motif detection in sequence segments (schematic).} 
The figure shows a configuration of correlated sequence sites of length $\ell=10$ bp and distance $r=14$ bp from each other. Pairs of correlated sites have the following properties: (i)~The average mutual similarity between aligned nucleotides is larger than a given threshold, $c \geq c_{\min}=0.8 $. (ii)~The left sites (and, hence, also the right sites) of all pairs have no common nucleotides. This condition is necessary in order to avoid overcounting of mutual similarity in overlapping site pairs. (iii)~The sum of the mutual similarities of all pairs in the set is maximal.  In the example shown, there are three different motifs with reoccurring sequence patterns marked by different colors (red, blue, green). To illustrate the alignment of the site pairs, we shift the whole sequence by $r=14$ bp in the second row. The left and right site of each motif are shown in boldface in the first and the  second row, respectively.  Mismatches between aligned sites of the same motif are shown in boldface gray letters. The flanking regions separating the correlated sequence pairs are shown in smaller font.}}
\end{figure}

\begin{figure}
\raggedright{}\includegraphics[scale=0.6]{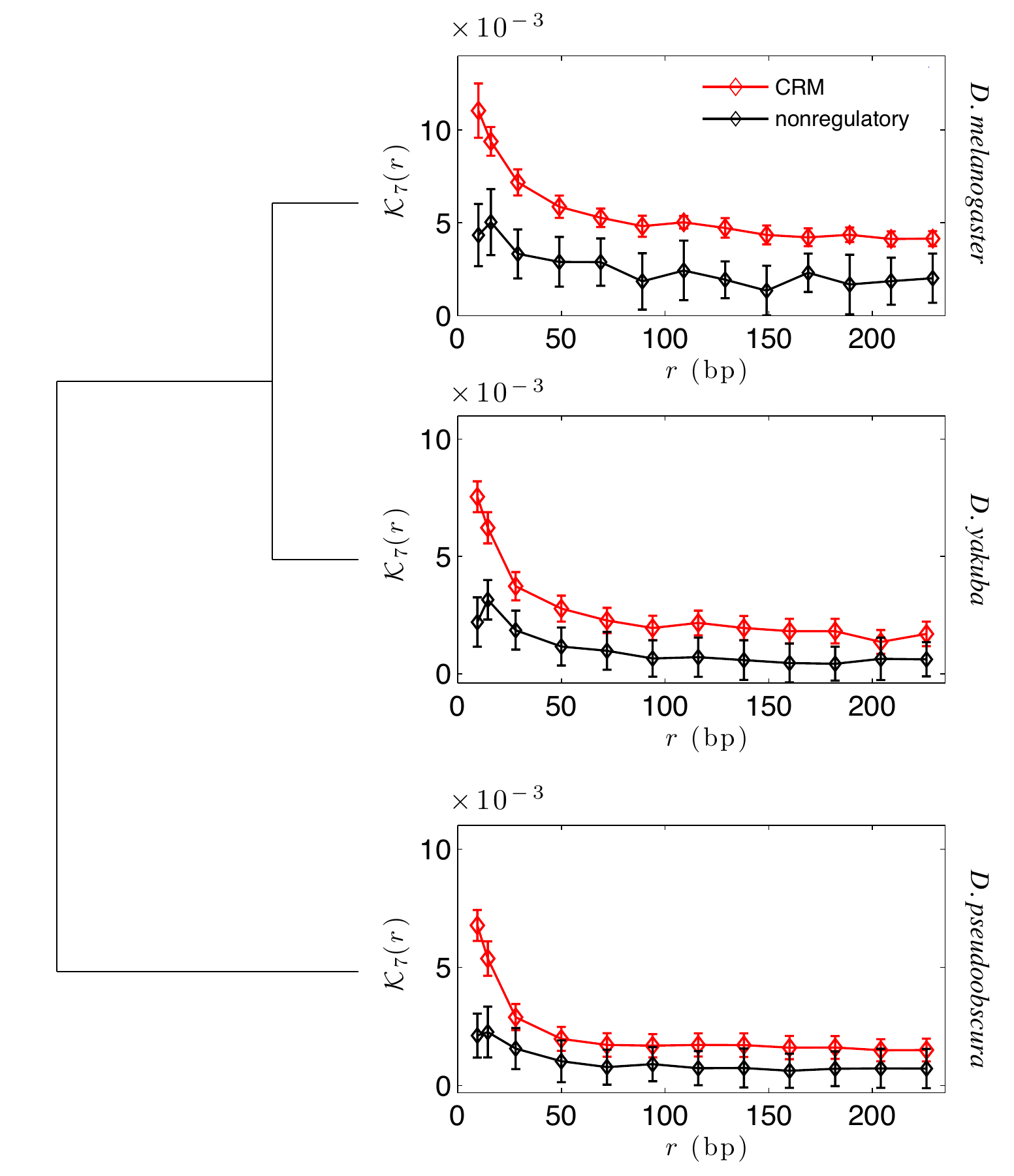}
\par
\caption{\label{fig:crossSpec} {\small {\bf Sequence similarity in regulatory modules of 3 {\em Drosophila} species.} 
Distance-dependent similarity information $K_7(r)$ for motif length $\ell = 7$ in regulatory modules (red) and in generic intergenic sequence (black), evaluated in  {\em D.~melangaster} and in the homologous regions of  {\em D.~ yakuba} and  {\em D.~pseudoobscura} (see Materials and Methods). These data show a consistent pattern of overall amplitudes and of decay lengths.  
}}
\end{figure}

\begin{figure}[T]
\raggedright{}\includegraphics[scale=0.43]{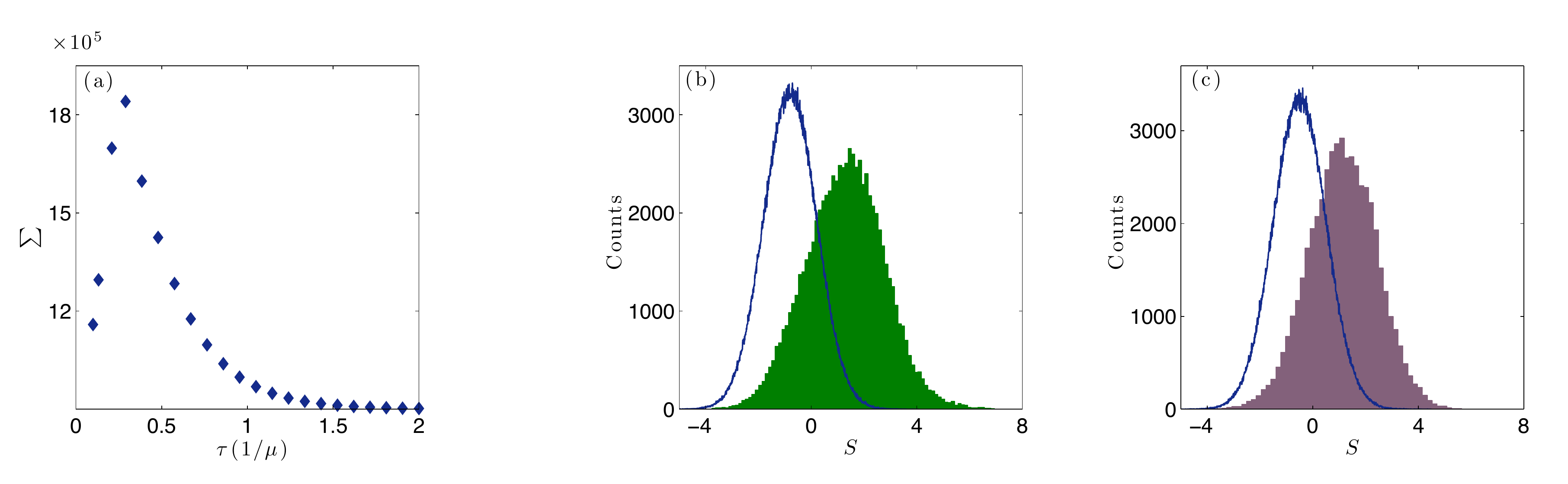}
\par
\caption{\label{fig:simulatedBS} {\small {\bf Tests of the duplication inference method.} 
We simulate binding site pairs $(\a,\b)$ evolving by common descent or by independent descent, as described in Materials and Methods. 
(a)~Dependence of the total duplication score $\Sigma$ on the time parameter $\tau$ for an ensemble of 150000 site pairs of common descent. This function has a pronounced maximum at a value $\tau_{\rm ML} \approx 0.3/\mu$, which is close to the mean divergence time $\bar \tau = 0.4/\mu$ since duplication. 
(b)~Distributions of the score $S$ (with $\tau = \tau_{\rm ML}$) for pairs of sites binding different factors. 
The distribution for sites of common descent (filled curve) is  distinguished from the distribution for sites with independent descent (solid curve) by its increased score average, $\langle S \rangle - \langle S \rangle_0 = 2.1$, and by its increased width. 
(c)~Same as (b) for pairs of sites binding the same factor. The distribution for sites of common descent (filled curve) has again an increased average, $\langle S \rangle - \langle S \rangle_0 = 1.6$, and an increased width. 
}}
\end{figure}

\end{document}